\documentclass[prl, nofootinbib, aps, twocolumn, superscriptaddress,
showpacs, preprintnumbers, amsmath, amssymb, floatfix]{revtex4}

\usepackage{graphicx}

\newcommand{\lsim}{\buildrel<\over{_\sim}}
\newcommand{\gsim}{\buildrel>\over{_\sim}}
\newcommand{\gtsim}{\gtrsim}

\newcommand{\GeV}{\mathrm{GeV}}
\newcommand{\TeV}{\mathrm{TeV}}
\newcommand{\Mpc}{\mathrm{Mpc}}
\newcommand{\km}{\mathrm{km}}
\newcommand{\seconds}{\mathrm{s}}

\newcommand{\MPl}{\mathrm{M}_{\mathrm{P}}}
\newcommand{\gravitino}{{\widetilde{G}}}
\newcommand{\stau}{{\widetilde{\tau}_1}}
\newcommand{\st}{{\tilde{\tau}_1}}

\newcommand{\mgr}{m_{\widetilde{G}}}
\newcommand{\mst}{m_{\tilde{\tau}_1}}

\newcommand{\NLSP}{\mathrm{NLSP}}
\newcommand{\NTP}{\mathrm{NTP}}
\newcommand{\TP}{\mathrm{TP}}

\newcommand{\CDM}{\mathrm{DM}}

\newcommand{\GUT}{\mathrm{GUT}}
\newcommand{\Reheating}{\mathrm{R}}
\newcommand{\Color}{\mathrm{c}}
\newcommand{\Weak}{\mathrm{L}}
\newcommand{\Hypercharge}{\mathrm{Y}}

\begin{document}
%
%
\preprint{hep-ph/0608344}
\preprint{MPP-2006-101}
%
%
%
\title{Thermal Gravitino Production and Collider Tests of Leptogenesis}
\author{Josef~Pradler}
\email{jpradler@mppmu.mpg.de}
\affiliation{Institut f\"ur Theoretische Physik, 
Universit\"at Wien, Boltzmanngasse 5, 
A--1090 Vienna, Austria 
 }
\affiliation{Max-Planck-Institut f\"ur Physik, 
F\"ohringer Ring 6,
D--80805 Munich, Germany}
\author{Frank Daniel Steffen}
\email{steffen@mppmu.mpg.de}
\affiliation{Max-Planck-Institut f\"ur Physik, 
F\"ohringer Ring 6,
D--80805 Munich, Germany}
%
%
%
\begin{abstract}
  Considering gravitino dark matter scenarios, we obtain the full
  gauge-invariant result for the relic density of thermally produced
  gravitinos to leading order in the Standard Model gauge couplings.
  For the temperatures required by thermal leptogenesis, we find
  gaugino mass bounds which will be probed at future colliders.
  We show that a conceivable determination of the gravitino mass will
  allow for a unique test of the viability of thermal leptogenesis in
  the laboratory.
\end{abstract}
\pacs{98.80.Cq, 95.35.+d, 12.60.Jv, 95.30.Cq}
%
%
\maketitle
\section{Introduction}

The smallness of the neutrino masses can be understood naturally in
terms of the see-saw mechanism~\cite{Minkowski:1977sc} once the
Standard Model is extended with right-handed neutrinos which have
heavy Majorana masses and only Yukawa couplings. For a reheating
temperature after inflation, $T_\Reheating$, which is larger or not
much smaller than the masses of the heavy neutrinos, these particles
are produced in thermal reactions in the early Universe. The
CP-violating out-of-equilibrium decays of the heavy neutrinos generate
a lepton asymmetry that is converted into a baryon asymmetry by
sphaleron processes~\cite{Fukugita:1986hr}. This mechanism, known as
thermal leptogenesis, can explain the cosmic baryon asymmetry for
$T_{\Reheating}\gsim 3\times 10^9\,\GeV$~\cite{Buchmuller:2004nz}.

One will face severe cosmological constraints on $T_\Reheating$ if
supersymmetry (SUSY) is discovered.  An unavoidable implication of
SUSY theories including gravity is the existence of the gravitino
$\gravitino$ which is the gauge field of local SUSY transformations.
As the spin-3/2 superpartner of the graviton, the gravitino is an
extremely weakly interacting particle with couplings suppressed by
inverse powers of the (reduced) Planck scale $\MPl=2.4\times
10^{18}\,\GeV$~\cite{Wess:1992cp}. In the course of spontaneous SUSY
breaking, the gravitino acquires a mass $\mgr$ and the couplings of
the spin-1/2 goldstino which become dominant for small $\mgr$.
Depending on the SUSY breaking scheme, $\mgr$ can range from the eV
scale up to scales beyond the TeV region~\cite{Martin:1997ns}.
Gravitinos can be produced efficiently in the hot primordial plasma.
Because of their extremely weak interactions, unstable gravitinos with
$\mgr \lsim 5~\TeV$ have long lifetimes, $\tau_{\gravitino} \gsim
100~\sec$, and decay after big bang nu\-cleo\-synthesis (BBN).  The
associated decay products affect the abundances of the primordial
light elements. Demanding that the successful BBN predictions are
preserved, bounds on the abundance of gravitinos before their decay
can be derived which imply $T_{\Reheating}\lsim 10^8\,\GeV$ for $\mgr
\lsim 5~\TeV$~\cite{Kohri:2005wn}. Thus, the temperatures needed for
thermal leptogenesis are excluded.

We therefore consider SUSY scenarios in which a gravitino with $\mgr
\gsim 10~\GeV$ is the lightest supersymmetric particle (LSP) and
stable due to R-parity conservation. These scenarios are particularly
attractive for two reasons: (i) the gravitino LSP is a compelling dark
matter candidate and (ii) thermal leptogenesis can still be a viable
explanation of the baryon asymmetry~\cite{Bolz:1998ek}.

\section{Thermal gravitino production}

Assuming that inflation governed the earliest moments of the Universe,
any initial population of gravitinos must be diluted away by the
exponential expansion during the slow-roll phase. We consider the
thermal production (or regeneration) of gravitinos in the
radiation-dominated epoch that starts after completion of reheating at
the temperature $T_{\Reheating}$.  Gravitinos with $\mgr \gsim
10~\GeV$ are not in thermal equilibrium with the primordial plasma
after inflation because of their extremely weak interactions.  At high
temperatures, gravitinos are generated in scattering processes of
particles that are in thermal equilibrium with the hot SUSY plasma.
The calculation of the relic density of these thermally produced
gravitinos, $\Omega_{\gravitino}^{\TP}$, requires a consistent
finite-temperature approach. A result that is independent of arbitrary
cutoffs has been derived for SUSY quantum chromodynamics (QCD) in a
gauge invariant way~\cite{Bolz:2000fu}. Following this approach, we
provide the complete
SU(3)$_\Color\times$SU(2)$_\Weak\times$U(1)$_\Hypercharge$ result to
leading order in the couplings.

We compute $\Omega_{\gravitino}^{\TP}$ from the Boltzmann equation
for the gravitino number density
\begin{equation}
    \frac{dn_{\gravitino}}{dt} + 3 H n_{\gravitino} = C_{\gravitino}
     \ .
\label{Eq:Boltzmann}
\end{equation}
The term proportional to the Hubble parameter $H$ accounts for the
dilution by the cosmic expansion.  The collision term $C_{\gravitino}$
describes the production and disappearance of gravitinos in thermal
reactions in the primordial plasma. Since the phase space density of
the gravitino is significantly below those of the particles in thermal
equilibrium, gravitino disappearance processes can be neglected. Thus,
$C_{\gravitino}$ is given by integrating the thermal gravitino
production rate.

Considering a primordial plasma with the particle content of the
minimal SUSY Standard Model (MSSM) in the high-temperature limit, we
calculate the thermal production rate of gravitinos with $E \gtsim T$
using the Braaten--Yuan prescription~\cite{Braaten:1991dd} and hard
thermal loop (HTL) resummation~\cite{Braaten:1989mz}. With the
systematic treatment of screening effects in the plasma, we obtain a
finite result in a gauge-invariant way.  Moreover, in contrast to
previous estimates~\cite{Ellis:1984eq,Moroi:1993mb}, our result does
not depend on arbitrary cutoffs. The explicit form of the thermal
gravitino production rate and the detailed derivation will be
presented in a forthcoming publication~\cite{Pradler:inPreparation}.

After numerical integration of the thermal gravitino production rate,
we obtain the
SU(3)$_\Color\times$SU(2)$_\Weak\times$U(1)$_\Hypercharge$ result for
the collision term:
\begin{equation}
        C_{\gravitino} 
        =
        \sum_{i=1}^{3} 
        \frac{3\zeta(3)T^6}{16\pi^3\MPl^2} 
        \left(1+\frac{M_i^2}{3\mgr^2}\right)
        c_i\, g_i^2
        \ln\left(\frac{k_i}{g_i}\right)
        \ ,
\label{Eq:CollisionTerm}        
\end{equation}
where the gaugino mass parameters $M_i$, the gauge couplings $g_i$,
and the constants $c_i$ and $k_i$ are associated with the gauge groups
U(1)$_\Hypercharge$, SU(2)$_\Weak$, and SU(3)$_\Color$ as given in
Table~\ref{Tab:Constants}.
\begin{table}[b]
  \caption{Assignments of the index $i$, the gauge coupling $g_i$, 
    and the gaugino mass parameter $M_i$ to the gauge groups
    U(1)$_\Hypercharge$, SU(2)$_\Weak$, and SU(3)$_\Color$
    and the values of the associated constants $c_i$, $k_i$, and $\omega_i$.}
  \label{Tab:Constants}
\begin{center}
\renewcommand{\arraystretch}{1.25}
\begin{tabular*}{3.25in}{@{\extracolsep\fill}ccccccc}
\hline\hline
gauge group         & $i$ & $g_i$ & $M_i$  & $c_i$ &  $k_i$ &  $\omega_i$
\\ \hline
U(1)$_\Hypercharge$ & 1 & $g'$    & $M_1$  & 11    & 1.266  & 0.018 
\\
SU(2)$_\Weak$       & 2 & $g$     & $M_2$  & 27    & 1.312  & 0.044 
\\
SU(3)$_\Color$ & 3 & $g_\mathrm{s}$ & $M_3$ & 72   & 1.271  & 0.117
\\
\hline\hline
\end{tabular*}
\end{center}
\end{table}
In expression~(\ref{Eq:CollisionTerm}) the temperature $T$ provides
the scale for the evaluation of $M_i$ and $g_i$.
Note that HTL resummation~\cite{Braaten:1989mz,Braaten:1991dd}
requires weak couplings, $g_i\ll 1$, and thus high temperatures $T\gg
10^6~\GeV$.

Our result $k_3=1.271$ for the SU(3)$_\Color$ contribution is larger
than $1.163$ obtained from~\cite{Bolz:2000fu}. This results from an
analytical disagreement~\cite{Pradler:inPreparation}: We find a
cancellation of the term
$T^3(N+n_f)[\mathrm{Li}_2(-e^{-E/T})-\pi^2/6]$ given as part of
$I_{BFB}$ in (C.14) of Ref.~\cite{Bolz:2000fu}.

Assuming conservation of entropy per comoving volume, the Boltzmann
equation~(\ref{Eq:Boltzmann}) can be solved
analytically~\cite{Bolz:2000fu,Brandenburg:2004du+X}. With the
collision term~(\ref{Eq:CollisionTerm}), we find
\begin{eqnarray}
        \Omega_{\gravitino}^{\TP}h^2
        &=&
        \sum_{i=1}^{3}
        \omega_i\, g_i^2 
        \left(1+\frac{M_i^2}{3\mgr^2}\right)
        \ln\left(\frac{k_i}{g_i}\right)
\nonumber\\
        &&
        \times
        \left(\frac{\mgr}{100~\GeV}\right)
        \left(\frac{T_{\Reheating}}{10^{10}\,\GeV}\right)
\label{Eq:GravitinoDensity}
\end{eqnarray}
with the Hubble constant $h$ in units of
$100~\km\,\Mpc^{-1}\seconds^{-1}$ and the constants $\omega_i$ given
in Table~\ref{Tab:Constants}.  Here $M_i$ and $g_i$ are understood to
be evaluated at $T_{\Reheating}$.
With our new $k_3$ value, we find an enhancement of about 30\% of the
SU(3)$_\Color$ contribution to the relic density.

\begin{figure}[t]
\begin{center}
\includegraphics[width=3.25in]{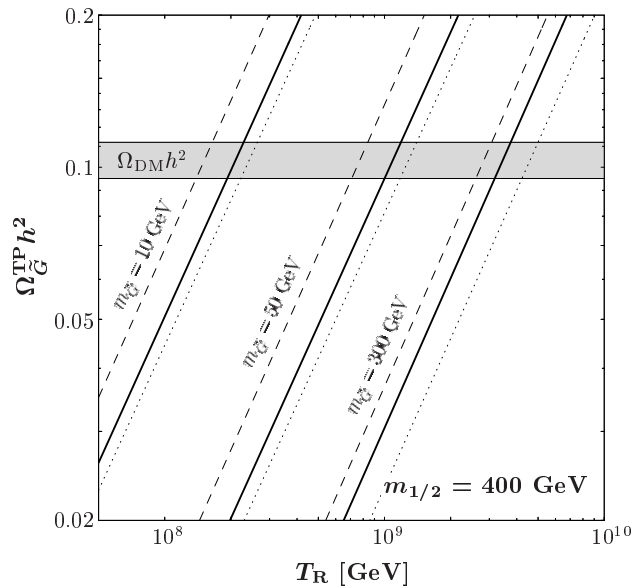} 
\caption{\small The relic gravitino density from thermal production,
  $\Omega_{\widetilde{G}}^{\TP}h^2$, as a function of
  $T_{\Reheating}$.  The solid and dashed curves show the
  SU(3)$_\Color\times$SU(2)$_\Weak\times$U(1)$_\Hypercharge$ results
  for universal ($M_{1,2,3}=m_{1/2}$) and non-universal ($0.5\,
  M_{1,2}=M_3=m_{1/2}$) gaugino masses at $M_{\GUT}$, respectively.
  The dotted curves show our result of the SU(3)$_\Color$ contribution
  for $M_3=m_{1/2}$ at $M_{\GUT}$. The gray band indicates the dark
  matter density $\Omega_{\CDM}h^2$.}
\label{Fig:Omegah2GravitinoTP}
\end{center}
\end{figure}

Figure~\ref{Fig:Omegah2GravitinoTP}
shows $\Omega_{\gravitino}^{\TP}h^2$ as a function of $T_{\Reheating}$
for $\mgr=10$, $50$, and $300~\GeV$. With $m_{1/2}=400~\GeV$, the
solid and dashed lines are obtained respectively for universal
($M_{1,2,3}=m_{1/2}$) and non-universal ($0.5\, M_{1,2}=M_3=m_{1/2}$)
gaugino masses at the grand unification scale $M_{\GUT} \simeq 2
\times 10^{16}\,\GeV$. The SU(3)$_\Color$ contributions are shown by
the dotted lines. The gray band indicates the dark matter
density~\cite{PDB2006}
\begin{equation}
        \Omega_{\CDM}h^2=0.105^{+0.007}_{-0.010}
        \ .
\label{Eq:OmegaCDMobs}
\end{equation}

We find that electroweak processes enhance $\Omega_{\gravitino}^{\TP}$
by about $20\%$ for universal gaugino masses at $M_{\GUT}$. In
non-universal cases, $M_{1,2}>M_3$ at $M_{\GUT}$, the electro\-weak
contributions are more important.  For $0.5\, M_{1,2}=M_3$ at
$M_{\GUT}$, they provide about $40\%$ of $\Omega_{\gravitino}^{\TP}$.

\section{Collider predictions of leptogenesis}

Thermal leptogenesis requires $T_{\Reheating}\gsim 3\times
10^9\,\GeV$~\cite{Buchmuller:2004nz}.  This condition together with
the constraint $\Omega_{\widetilde{G}}^{\TP}\leq\Omega_{\CDM}$ leads
to upper limits on the gaugino masses.  The SU(3)$_\Color$ result for
$\Omega_{\widetilde{G}}^{\TP}$ implies limits on the gluino
mass~\cite{Bolz:2000fu,Fujii:2003nr}.  With our
SU(3)$_\Color\times$SU(2)$_\Weak\times$U(1)$_\Hypercharge$ result, the
limits on the gluino mass $M_3$ become more stringent because of the
new $k_3$ value and the additional electroweak contributions.
Moreover, as a prediction of thermal leptogenesis, we obtain upper
limits on the electro\-weak gaugino mass parameters $M_{1,2}$.  At the
Large Hadron Collider (LHC) and the International Linear Collider
(ILC), these limits will be probed in measurements of the masses of
the neutralinos and charginos, which are typically lighter than the
gluino. If the superparticle spectrum does not respect these bounds,
one will be able to exclude standard thermal leptogenesis.

\begin{figure}[t]
\begin{center}
\includegraphics[width=3.25in]{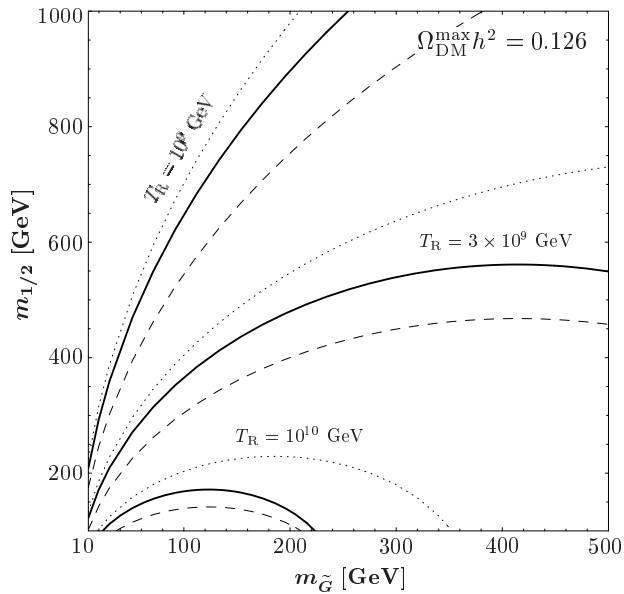} 
\caption{\small Upper limits on the gaugino mass parameter $m_{1/2}$
  from $\Omega_{\widetilde{G}}^{\TP}\leq\Omega_{\CDM}^{\max}$ for the
  indicated values of $T_{\Reheating}$. The solid and dashed curves
  show our SU(3)$_\Color\times$SU(2)$_\Weak\times$U(1)$_\Hypercharge$
  results for universal ($M_{1,2,3}=m_{1/2}$) and non-universal
  ($0.5\, M_{1,2}=M_3=m_{1/2}$) gaugino masses at $M_{\GUT}$,
  respectively.  The dotted curves show the SU(3)$_\Color$ limits for
  $M_3=m_{1/2}$ at $M_{\GUT}$.}
\label{Fig:UpperLimitM12}
\end{center}
\end{figure}

Figure~\ref{Fig:UpperLimitM12}
shows the gaugino mass bounds for $T_{\Reheating}=10^9$, $3\times
10^9$, and $10^{10}~\GeV$ evolved to $M_{\GUT}$, i.e., in terms of
limits on the gaugino mass parameter $m_{1/2}$.
Here $\Omega_{\CDM}^{\max}h^2=0.126$ is adopted as a nominal $3\sigma$
upper limit on $\Omega_{\CDM}h^2$.
With the observed superparticle spectrum, one will be able to evaluate
the gaugino mass parameters $M_{1,2,3}$ at $M_{\GUT}$ using the SUSY
renormalization group
equations~\cite{Baer:2000gf,Blair:2002pg,Lafaye:2004cn+X}.  While the
determination of $M_{1,2}$ at low energies depends on details of the
SUSY model that will be probed at colliders~\cite{Martin:1997ns}, the
bounds shown in Fig.~\ref{Fig:UpperLimitM12} depend mainly on the
$M_i$ relation at $M_{\GUT}$. This is illustrated by the solid and
dashed curves obtained with $M_{1,2,3}=m_{1/2}$ and $0.5\,
M_{1,2}=M_3=m_{1/2}$, respectively. The dotted curves represent the
SU(3)$_\Color$ limits for $M_3=m_{1/2}$ at $M_{\GUT}$ and emphasize
the importance of the electroweak contributions.

\section{Decays of the next-to-lightest supersymmetric particle}

With a gravitino LSP of $\mgr \gsim 10~\GeV$, the next-to-lightest
SUSY particle (NLSP) has a long lifetime of $\tau_{\NLSP}\gsim
10^6\,\seconds$~\cite{Feng:2004mt,Steffen:2006hw}.  After decoupling
from the primordial plasma, each NLSP decays into one gravitino LSP
and Standard Model particles. The resulting relic density of these
non-thermally produced gravitinos is
\begin{equation}
        \Omega_{\widetilde{G}}^{\NTP}h^2
        = \frac{\mgr}{m_{\NLSP}}\,\Omega_{\NLSP}h^2
        \ ,
\label{Eq:GravitinoDensityNTP}
\end{equation}
where $m_{\NLSP}$ is the mass of the NLSP and $\Omega_{\NLSP}h^2$ is
the relic density that the NLSP would have today, if it had not
decayed. As shown below, more severe limits on $m_{1/2}$ are obtained
with $\Omega_{\widetilde{G}}^{\NTP}h^2$ taken into account.  Moreover,
since the NLSP decays take place after BBN, the emitted Standard Model
particles can affect the abundance of the primordial light elements.
Successful BBN predictions thus imply bounds on $\mgr$ and
$m_{\NLSP}$~\cite{Feng:2004mt,Steffen:2006hw}. From these cosmological
constraints it has been found that thermal leptogenesis remains viable
only in the cases of a charged slepton NLSP or a sneutrino
NLSP~\cite{Fujii:2003nr,Cerdeno:2005eu}.

Note that the cosmological constraints from BBN can become much weaker
with entropy production after decoupling of the NLSP and before BBN.
For example, large parts of the parameter region disfavored by BBN
constraints on charged slepton NLSP scenarios become allowed with a
moderate amount of entropy production~\cite{Buchmuller:2006tt}.  In
addition, such an entropy production dilutes both the generated baryon
asymmetry and the thermally produced abundance of gravitinos.
Therefore, upper limits on the gaugino masses will still allow us to
probe the viability of thermal leptogenesis.  However, the
$T_{\Reheating}$ labels of the curves in Fig.~\ref{Fig:UpperLimitM12}
will change to higher values. 

\section{Collider Tests of Leptogenesis}

Thermal leptogenesis will predict a lower bound on the gravitino mass
$\mgr$ once the masses of the Standard Model superpartners are known.
With a charged slepton as the lightest Standard Model superpartner, it
could even be possible to identify the gravitino as the LSP and to
measure its mass $\mgr$ at future
colliders~\cite{Buchmuller:2004rq,Brandenburg:2005he+X,Martyn:2006as}.
Confronting the measured $\mgr$ with the predicted lower bound will
then allow us to decide about the viability of thermal leptogenesis.

In order to explain our method, we do now consider an exemplary SUSY
model. Let us assume that the analysis of the observed
spectrum~\cite{Lafaye:2004cn+X} will point to the universality of the
soft SUSY breaking parameters at $M_{\GUT}$ and, in particular, to the
minimal supergravity (mSUGRA) scenario with the gaugino mass parameter
$m_{1/2}=400~\GeV$, the scalar mass parameter $m_0=150~\GeV$, the
trilinear coupling $A_0=-150~\GeV$, a positive higgsino mass
parameter, $\mu>0$, and the mixing angle $\tan\beta=30$ in the Higgs
sector.
A striking feature of the spectrum will then be the appearance of the
lighter stau $\stau$ with $\mst=143.4~\GeV$ as the lightest Standard
Model superpartner~\cite{Djouadi:2002ze}.
In the considered gravitino LSP case, $10~\GeV\lsim\mgr<\mst$, this
stau is the NLSP and decays with a lifetime of
$\tau_{\st}\gsim 10^6\,\seconds$
into the gravitino.
For the identified mSUGRA scenario and the considered reheating
temperatures, the cosmological abundance of the $\stau$ NLSP prior to
decay can be computed from 
$\Omega_{\NLSP}h^2=\Omega_{\stau}h^2\simeq 3.83\times 10^{-3}$,
which is provided by the computer program
micrOMEGAs~\cite{Belanger:2001fz+X}.
For given $\mgr$, this abundance determines
$\Omega_{\gravitino}^{\NTP}h^2$ and the release of electromagnetic
(EM) and hadronic energy in $\stau$ NLSP decays governing the
cosmological constraints~\cite{Feng:2004mt,Steffen:2006hw}.\footnote{
  Considering bound-state effects of long-lived negatively charged
  particles on BBN, it has recently been claimed that the stau NLSP
  abundance for~$\tau_{\st}\gsim 10^3-10^4\,\seconds$ is severely
  constrained by the observed $^{6,7}$Li
  abundances~\cite{Pospelov:2006sc,Cyburt:2006uv}. If the associated
  bounds are confirmed, the considered mSUGRA scenario will be
  cosmologically allowed only with late-time entropy
  production~\cite{Buchmuller:2006tt}.}

\begin{figure}[t]
\begin{center}
\includegraphics[width=3.25in]{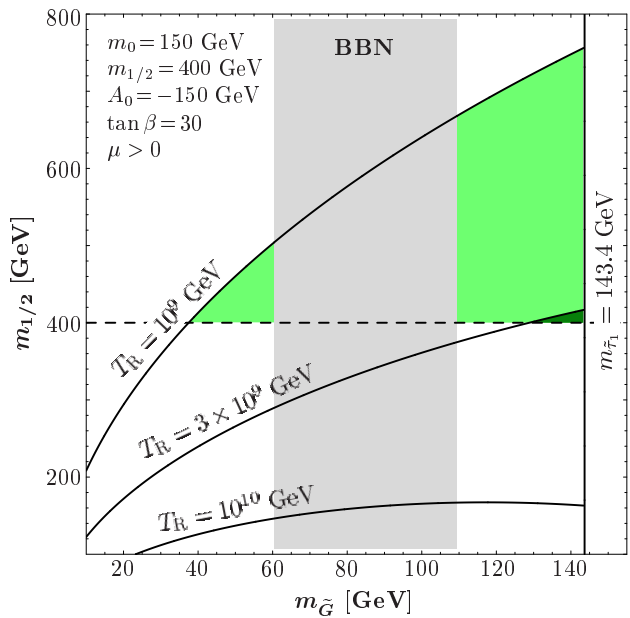} 
\caption{\small Probing the viability of thermal leptogenesis. The
  solid curves show the limits on the gaugino mass parameter $m_{1/2}$
  from
  $\Omega_{\widetilde{G}}^{\TP}+\Omega_{\widetilde{G}}^{\NTP}\leq\Omega_{\CDM}^{\max}$
  for $T_{\Reheating}=10^9$, $3\times 10^9$, and $10^{10}~\GeV$.  The
  dashed line indicates the $m_{1/2}$ value of the considered
  scenario. The vertical solid line is given by the $\stau$ NLSP mass
  which limits the gravitino LSP mass from above:
  $\mgr<\mst=143.4~\GeV$. The $\mgr$ values at which temperatures
  above $3\times 10^9~\GeV$ and $10^9~\GeV$ remain allowed are
  indicated by the dark-shaded (dark-green) and medium-shaded
  (light-green) regions, respectively.  The $\mgr$ values within the
  light-shaded region are excluded by BBN constraints.}
\label{Fig:ProbingTLGViability}
\end{center}
\end{figure}

Figure~\ref{Fig:ProbingTLGViability}
allows us to probe the viability of thermal leptogenesis in the
considered mSUGRA scenario.\footnote{Thermal leptogenesis requires
  right-handed neutrinos and thus an extended mSUGRA scenario. This
  could ma\-ni\-fest itself in the masses of the third generation
  sleptons~\cite{Baer:2000gf}. Since the effects are typically small,
  we leave a systematic investigation of extended scenarios for future
  work.} 
From the constraint
$\Omega_{\widetilde{G}}^{\TP}+\Omega_{\widetilde{G}}^{\NTP}\leq\Omega_{\CDM}^{\max}$,
we obtain the solid curves which provide the upper limits on $m_{1/2}$
for $T_{\Reheating}=10^9$, $3\times 10^9$, and $10^{10}~\GeV$.
The dashed line indicates the $m_{1/2}$ value of the considered
scenario. The vertical solid line is given by $\mst=143.4~\GeV$ which
limits $\mgr$ from above.
In the considered scenario, the $m_{1/2}$ value exceeds the $m_{1/2}$
limits for $T_{\Reheating}\gsim 10^{10}~\GeV$. Thus, temperatures
above $10^{10}~\GeV$ can be excluded.
Temperatures above $3\times 10^9~\GeV$ and $10^9~\GeV$ remain allowed
for $\mgr$ values indicated by the dark-shaded (dark-green) and
medium-shaded (light-green) regions, respectively.
The $\mgr$ values indicated by the light-shaded region are excluded by
BBN constraints for late $\stau$ NLSP decays.\footnote{We use the
  conservative BBN bounds considered in~\cite{Steffen:2006hw}.  The
  average EM energy release in one $\stau$ NLSP decay is assumed to be
  $E_{\tau}/2$, where $E_{\tau}$ is the energy of the tau emitted in
  the dominant 2-body decay $\stau\to\gravitino\tau$ (cf.\ Fig.~16 of
  Ref.~\cite{Steffen:2006hw}). With an EM energy release below
  $E_{\tau}/2$, the light-shaded band can become smaller. For less
  conservative BBN constraints and/or enhanced EM energy release, the
  excluded $\mgr$ region becomes larger.}

Here thermal leptogenesis, $T_{\Reheating}\gsim 3\times 10^9\,\GeV$,
predicts $\mgr\gsim 130~\GeV$ and thus a $\stau$ lifetime of
$\tau_{\st}>10^{11}~\seconds$~\cite{Feng:2004mt,Steffen:2006hw}.  If
decays of long-lived $\stau$'s can be analyzed at colliders giving
evidence for the gravitino
LSP~\cite{Buchmuller:2004rq,Brandenburg:2005he+X,Martyn:2006as}, there
will be the possibility to determine $\mgr$ in the laboratory: From a
measurement of the lifetime $\tau_{\st}$ governed by the decay
$\stau\to\gravitino\tau$, $\mgr$ can be extracted using the
supergravity prediction for the associated partial width,
\begin{equation}
        \tau_{\st} 
        \simeq \Gamma^{-1}(\stau\to\gravitino\tau)
        = \frac{48 \pi \mgr^2 \MPl^2}{m_{\st}^5} 
        \left(1-\frac{\mgr^2}{m_{\st}^2}\right)^{-4}
\label{Eq:StauLifetime}
\end{equation}
as obtained for $m_{\tau}\to 0$.
Moreover, for $\mgr\gsim 0.1\,\mst$, $\mgr$ can be infered
kinematically from the energy of the tau, $E_{\tau}$, emitted in the
2-body decay
$\stau\to\gravitino\tau$~\cite{Buchmuller:2004rq,Martyn:2006as}:
\begin{equation}
        \mgr = \sqrt{m_{\st}^2-m_{\tau}^2-2 m_{\st} E_{\tau}}
        \ . 
\label{Eq:GravitinoMassKinematically}
\end{equation}
While $\mgr$ within the dark-shaded (dark-green) region will favor
thermal leptogenesis, any $\mgr$ outside of the medium-shaded
(light-green) region will require either non-standard mechanisms
lowering the $T_{\Reheating}$ value needed for thermal leptogenesis or
an alternative explanation of the cosmic baryon asymmetry.

\section{Conclusion}

We provide the full
SU(3)$_\Color\times$SU(2)$_\Weak\times$U(1)$_\Hypercharge$ result for
the relic density of thermally produced gravitino LSPs to leading
order in the gauge couplings.
Our result is obtained in a consistent gauge-invariant
finite-temperature calculation and thus independent of arbitrary
cutoffs.
With this result, new gravitino and gaugino mass bounds emerge as a
prediction of thermal leptogenesis.
If supersymmetry is realized in Nature, these bounds will be
accessible at the LHC and the ILC.
In particular, with a charged slepton NLSP, there will be the exciting
possibility to identify the gravitino as the LSP and to measure its
mass.
Confronting the measured gravitino mass with the predicted bounds will
then allow for a unique test of the viability of thermal leptogenesis
in the laboratory.

\medskip

\begin{acknowledgments}
  We are grateful to T.~Plehn, M.~Pl\"umacher, G.~Raffelt, and
  Y.Y.Y.~Wong for valuable discussions.
\end{acknowledgments}
%

%
%
\end{document}